\documentclass[a4paper,11pt]{article}
\usepackage{pos}
\usepackage[T1]{fontenc}
\usepackage{graphicx}
\usepackage{amsmath}
\usepackage{amsfonts}
\usepackage{latexsym}
\usepackage{slashed}
\usepackage{hyperref}
\usepackage{feynmp-auto}
\usepackage{mathtools}
\usepackage{subcaption}
\usepackage{tabularx}
\usepackage{graphicx}

\begin{document}
\title{Measuring neutrino dynamics in NMSSM with a right-handed sneutrino LSP at the ILC}
\author*[a]{Yi Liu}
\author[a,b,c]{Stefano Moretti}
\author[a,b,c]{Harri Waltari}

\affiliation[a]{School of Physics and Astronomy, University of Southampton, \\
Highfield, Southampton SO17 1BJ, UK}
\affiliation[b]{Particle Physics Department, STFC Rutherford Appleton Laboratory, \\
Chilton, Didcot, Oxon OX11 0QX, UK}
\affiliation[c]{Department of Physics and Astronomy, Uppsala University, \\
Box 516, S-75120 Uppsala, Sweden}

\emailAdd{Yi.Liu@soton.ac.uk}
\emailAdd{S.Moretti@soton.ac.uk}
\emailAdd{H.Waltari@soton.ac.uk}

\abstract{We study the possibility of measuring neutrino Yukawa couplings in the Next-to-Minimal Supersymmetric Standard Model  with right-handed neutrinos (NMSSMr) when the lightest Supersymmetric partner (a right-handed sneutrino) is the Dark Matter (DM) candidate, by exploiting a `dijet + dilepton + Missing Transverse Energy' (MET)  signature. We show that, unlike the miminal realisation of Supersymmetry (SUSY), i.e., the MSSM, for which the DM candidate is  a much heavier (fermionic)  state (a neutralino), this non-minimal SUSY model offers one with a much lighter (bosonic) state (a sneutrino) as DM, which can be produced at future $e^+e^-$ colliders with energies up to about 500 GeV. The emerging signal, from chargino pair production and subsequent decay, is extremely pure so it allows for the possibility of extracting the Yukawa parameters of the (s)neutrino sector. These results should then motivate searches for light DM signals at such accelerators, where a DM candidate with mass at the  Electro-Weak (EW) scale can then be accessed.}

\FullConference{41st International Conference on High Energy physics - ICHEP2022\\6-13 July, 2022\\Bologna, Italy}

\maketitle

\section{Introduction}

The Large Hadron Collider (LHC) experiments have so far shown good agreement with the predictions of the Standard Model (SM). Other types of experiments instead show us that the SM needs to be extended: e.g.,  neutrino oscillations require neutrinos to be massive \cite{Athanassopoulos:1997pv}. A seesaw mechanism is a natural way to explain neutrino masses \cite{Minkowski:1977sc}. Also the Cosmic Microwave Background (CMB) \cite{Ade:2013zuv} and galactic rotation curves \cite{Zwicky:1933gu} strongly support the idea that most of the mass of the Universe is in a form currently unknown to us, dubbed DM. SUSY, as one of the most studied frameworks to construct Beyond the SM (BSM) theories, offers a natural DM candidate. The combination of SUSY and seesaw mechanism gives an option of non-standard DM candidates. Specifically, the right-handed sneutrino, when is the Lightest Supersymmetric Particle (LSP), has been of considerable interest over the years \cite{Moretti:2019ulc, Chakraborty:2020cpa, Asaka:2005cn}. In the MSSM, extended with Type-I seesaw right-handed sneutrinos, one finds an  overabundance of the relic density of the CMB, unless there is significant mixing between the left- and right-handed sneutrinos \cite{ArkaniHamed:2000bq}. Adding a singlet to the model, ({i.e.}, considering the NMSSMr, allows a coupling between the heavy Higgses and sneutrinos, which can assist in the DM annihilation and lead to the correct relic abundance without the need for any left-right mixing in the sneutrino sector \cite{Cerdeno:2008ep}.

\section{Model description and signal vs background definition}

The NMSSM extends the MSSM with an additional gauge singlet chiral superfield $S$. The NMSSM fixes the $\mu$ problem of the MSSM by generating an effective $\mu$-term, but  it still inherits the defect that neutrinos are massless. By adding a singlet right-handed neutrino superfield, $N$, one may introduce the Type-I seesaw mechanism to generate neutrino masses. The Superpotential is given by \cite{Cerdeno:2008ep}

\begin{equation}
        W = W_{\rm NMSSM} + \lambda_N SNN + y_N H_2\cdot LN,
        \label{1}
    \end{equation}
    \begin{equation}
        W_{\rm NMSSM} = y_uH_2\cdot Qu + y_d H_1\cdot Qd + y_e H_1\cdot Le - \lambda SH_1\cdot H_2 + \frac{1}{3}\kappa S^3.
        \label{2}
    \end{equation} 

\begin{figure}
    \centering
    \includegraphics[scale=0.4]{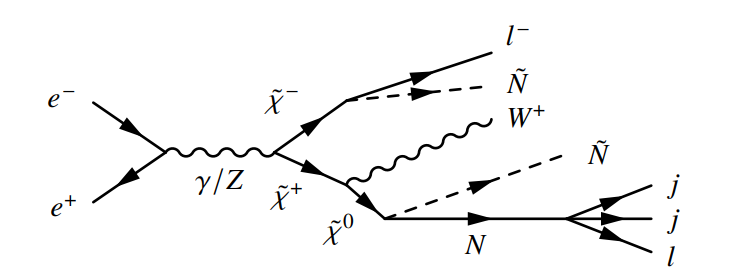}
    \caption{An example of a full process leading to the `dijet + dilepton + $\slashed{E}_T$' signature.}
    \label{signal}
\end{figure}

We consider the process $e^+ e^-\to \gamma^{*} /Z^{*} \to \tilde{\chi}^+\tilde{\chi}^-$ with one of the charginos decaying to a lepton and a sneutrino and the other  decaying to a neutralino and a virtual $W^\pm$ leading to a soft lepton plus MET or to hadrons. If the neutralino decays invisibly through the neutrino Yukawa couplings, {i.e.}, via $\tilde{\chi}^{0}\rightarrow \tilde{N}\nu$, the final state with a single hard lepton and MET will get a too large background from $W^\pm$ bosons. If, however, the right-handed neutrino and sneutrino are light enough, the neutralino can decay via its singlino component through $\tilde{\chi}^{0}\rightarrow\tilde{N}N$ with the right-handed neutrino decaying subsequently to a lepton and two jets. The Feynman diagram of this process is shown in Figure \ref{signal}. Requiring a lepton and two jets with an invariant mass correspoding to the right-handed neutrino mass will be enough to get rid of the backgrounds, as we shall see. Finally, notice that the existence of right-handed neutrinos can be established already at the LHC.

We can then get the aforementioned signature: `dijet + dilepton + MET'. The dilepton component should emerge in both same-sign and opposite-sign dileptons due to the Majorana nature of the right-handed neutrino. The latter will have a smaller background from SM processes. The major SM background to this final state comes from the following processes.

\begin{itemize}
    \item $W^+W^-Z$ production in the case where one $W^\pm$ boson decays into two jets and the other to a lepton and neutrino while the $Z$ boson gives two leptons, one of which is missed by the detector.
    \item $ZZZ$ production, where the first $Z$ boson decays leptonically, the second to neutrinos and the third creates the two jets.
    \item $t\bar{t}$ production, where the top (anti)quarks decay to a $W^\pm$ boson and a $b$-quark, when one lepton originates from the $W^\pm$ boson and another lepton from a $B$-meson.
\end{itemize}
We now proceed to describe how we performed our Monte Carlo (MC) analysis.

\section{Event simulation}

We prepared a number of Benchmark Points (BPs), which can potentially be probed at the $\sqrt{s}=500$~GeV phase of the International Linear Collider (ILC), for which we use an integrated luminosity of $4000$~fb$^{-1}$ \cite{Barklow:2015tja}. We select the charginos to be slightly lighter than $250$~GeV and the right-handed neutrino and sneutrino so light that $\tilde{\chi}^{0}\rightarrow \tilde{N}N$ is kinematically allowed. We show the spectra {and  relevant Yukawa couplings} of our BPs in Table  \ref{tb:benchmarks}. We checked with \textsc{MadDM} v3.0 \cite{Ambrogi:2018jqj} that  the BPs are acceptable with respect to constraints from the relic density and direct detetion experiments. Regarding the relic density we only imposed the upper limit $\Omega h^{2}\leq 0.12$.

\begin{table}[h]
\begin{center}
\begin{tabular}{|c|c|c|c|}
\hline
 & BP1 & BP2 & BP3\\
\hline
$m(\tilde{\chi}_{1}^{\pm})$ (GeV) & 239.3 & 234.8 & 233.3\\
$m(\tilde{\chi}_{1}^{0})$ (GeV) & 233.3 & 228.7 & 227.3\\
$m(\tilde{N}_{1})$ (GeV) & 130.6 & 127.9 & 127.4\\
$m(N_{1})$ (GeV) & 101.7 & 90.5 & 88.6\\
$\mathrm{BR}(N\rightarrow \ell jj)$ & $60\%$ & $68\%$ & $68\%$ \\
$\mathrm{BR}(W^{*}\rightarrow \mathrm{leptons})$ & $28\%$ & $28\%$ & $28\%$\\
$\tan \beta$ & $2.3$ & $2.4$ & $2.1$\\
$y^{\nu}_{1j}$, $y^{\nu}_{2j}$ ($10^{-7}$) & $5.3$, $3.5$ & $6.1$, $4.7$  & $5.3$, $4.0$\\
\hline
\end{tabular}
\end{center}
\caption{Mass spectra and other relevant parameters of our BPs. {The neutrino Yukawa couplings are given for the flavour of the lightest right-handed sneutrino. The amplitude of our signal process will be proportional to these Yukawa couplings}.\label{tb:benchmarks}}
\end{table}

As a preselection we require two same-sign leptons, two jets and veto against $b$-jets. For the $b$-tagger we use as working assumption of   $b$-tagging efficiency a value of $70\%$ with a mistagging rate of $2\%$ for $c$-quark jets and $0.3\%$ for light quark (and gluon) jets. We impose the preselection and full cuts given in Table \ref{cutresult}, where the final signal and background rates can also be found.

\begin{figure}
    \centering
    \includegraphics[scale=0.3]{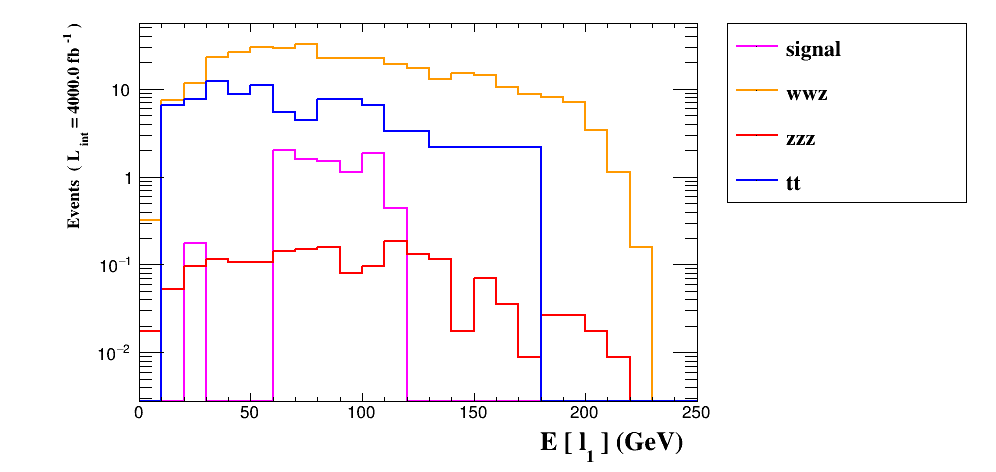}
    \caption{The energy of the leading lepton $\ell_1$ for our signal and different background components.}
    \label{fig:el1}
    \hfill
\end{figure}

\begin{figure}
    \centering
    \includegraphics[scale=0.3]{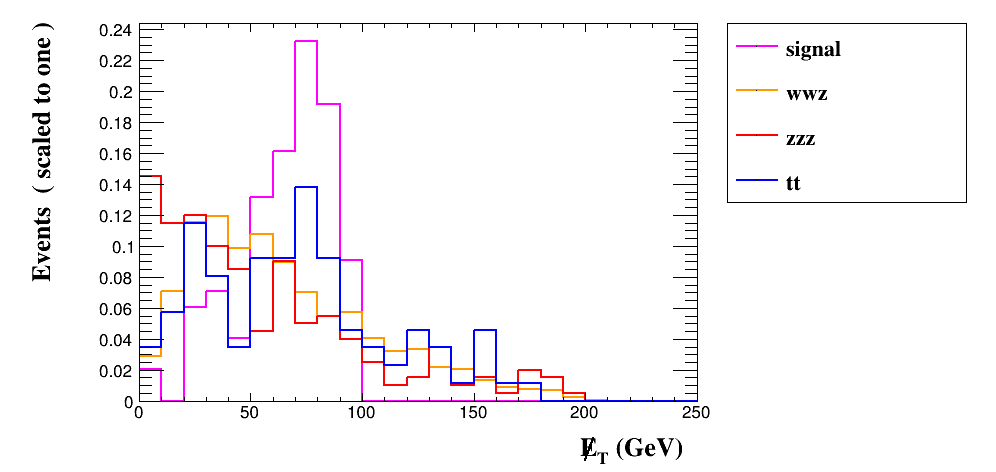}
    \caption{The MET distribution  for the signal and background components. Here we have normalised the distributions to unity.}
    \label{MET}
\end{figure}

\begin{table}[]
    \centering
\hspace*{-1.25truecm}
    \begin{tabular}{|c|c c c|c c c c|}
    \hline
    Cut & BP1 & BP2 & BP3 & $W^+W^-Z$ & $ZZZ$ & $t\bar t$ & Total background \\
    \hline
    Initial & 87.0 & 139 & 116 & 158999 & 4400 & 2193599 & 2356998 \\
    $b$-jet veto & 84.2 & 137 & 115 & 133754 & 2802 & 240648 & 377204 \\
    $N(\ell)$=2 & 38.8 & 54.9 & 42.0 & 11308 & 387 & 11454 & 23149 \\
    $N(\ell^+)=2$ or $N(\ell^-)=2$ & 17.8 & 26.0 & 20.6 & 792 & 6.07 &339 & 1137 \\
    $N(j)=2$ & 8.66 & 12.3 & 8.69 & 343 & 1.76 & 95.4 & 440\\
    \hline
   
    $p_T(j_1)<70$ GeV & 8.66 & 12.0 & 8.35 & 154.5 & 0.625 & 26.3 & 181.4 \\
    $p_T(\ell_1)>30$ GeV & 7.87 & 10.2 & 8.11 & 134.5 & 0.519 & 17.6 & 152.6 \\
    $p_T(\ell_2)<40$ GeV & 7.87 & 10.2 & 8.11 & 95.7 & 0.36 & 17.6 & 113.7 \\
    $H_T<100$ GeV & 7.87 & 10.2 & 8.00 & 76.5 & 0.24 & 11.0 & 87.7 \\
    
    $E(\ell_1)<120$ GeV & 7.87 & 10.2 & 8.00 & 55.5 & 0.176 & 7.68 & 63.4 \\
    $E(\ell_1)>60$ GeV & 7.87 & 9.33 & 7.65 & 36.6 & 0.123 & 5.48 & 42.2 \\
    $\Delta\Phi_{0,\pi}>$2.5 & 7.70 & 8.08 & 6.14 & 16.7 & 0.035 & 3.29 & 20.0 \\
    ${\rm MET}>50$ GeV & 6.82 & 7.38 & 4.98 & 9.70 & 0.026 & 2.19 & 11.9 \\
    ${\rm MET}<100$ GeV & 6.82 & 5.99 & 4.06 & 8.27 & 0.026 & 2.19 & 10.5 \\

    $M(\ell_1\ell_2)<80$ GeV & 5.60 & 5.71 & 3.94 & 4.77 & 0.018 & 1.10 & 5.89 \\
    $M(j_1j_2\ell_2)<110(100)$ GeV & 5.51 & 5.71 & 3.94 & 2.23(1.40) & 0.0088(0) & 1.10(1.10) & 3.34(2.50) \\
    $M(j_1j_2\ell_2)>90(80)$ GeV & 3.67 & 3.48 & 2.43 & 1.11(0.636) & 0.0088(0) & 0(0) & 1.1(0.64)\\
    \hline
    \end{tabular}
    \caption{The cutflow for the signal BPs and all background. The luminosity is 4000 fb$^{-1}$ and the energy is $\sqrt{s}=500$~GeV. The bracket stands for the cut and result corresponding to both BP2 and BP3. After these cuts, the significance for BP1 is $3.5\sigma$, for BP2 is $4.4\sigma$ and for BP3 is $3.0\sigma$.}
    \label{cutresult}
\end{table}

Overall, we are able to see a significant excess above the expected SM background, showing evidence of the neutrino mass generation mechanism, but a full discovery would need a higher integrated luminosity. Nevertheless, even if the excesses are not statistically significant enough to claim the discovery of the two-body decay of the chargino, some bounds on the neutrino Yukawa couplings can be inferred.

\section{Estimating neutrino Yukawa couplings}

The coupling between the right-handed sneutrino, charged lepton and lightest chargino is
\begin{equation}
\lambda_{\tilde{N}\ell^{+}\tilde{\chi}^{-}}=\frac{i}{\sqrt{2}}y^{\nu}_{ab}V_{12}\frac{1+\gamma_{5}}{2},
\end{equation}
where $a,b$ refer to neutrino flavours and $V_{12}$ gives the higgsino component of the lightest chargino. For our BPs, we have $|V_{12}|\simeq 1$.
This leads to the following decay width (neglecting the lepton mass):
\begin{equation}\label{eq:charginowidth1}
\Gamma(\tilde{\chi}^{\pm}\rightarrow \ell^{\pm}_{a}\tilde{N}_{b})=\frac{(m_{\tilde{\chi}}^{2}-m_{\tilde{N}}^{2})^{2}}{64\pi m_{\tilde{\chi}}^{3}}|y^{\nu}_{ab}|^{2}|V_{12}|^{2}.
\end{equation}
The majority of charginos decay via $\tilde{\chi}^{\pm}\rightarrow \tilde{\chi}^{0}\ell^{\pm}\nu,  \tilde{\chi}^{0}q\overline{q}^{\prime}$. This decay can in principle be mediated by several particles ($W^{\pm}$, $\tilde{\ell}^{\pm}$, $\tilde{\nu}$, $H^{\pm}$, $\tilde{q}$) and their contributions can interfere. The measurement of the Branching Ratio (BR) of this rare chargino decay would give us an estimate of the neutrino Yukawa couplings through the computed full width. The calculation details are shown in \cite{Yi:2021}.

\section{Conclusions}

In this paper, 
we have shown that, if the right-handed neutrino of the NMSSMr is the DM candidate, it can be so light that the neutral higgsino will decay via $\tilde{\chi}^{0}\rightarrow N\tilde{N}$, so that this additional handle will let us find the rare two-body decay of the chargino,  $\tilde{\chi}^-\to l^-\tilde N$, at the ILC while at the LHC such a discovery is impossible as the boost between the laboratory  and  CM frames is unknown. This decay would then allow us to estimate the size of the neutrino Yukawa couplings but, even with the luminosity upgrade of the ILC, the measurement will be statistically limited and give at best an accuracy of $25\%$. Even this limited accuracy should be enough, though, to test the consistency of the underlying (s)neutrino mass model, {i.e.}, that the neutrino masses are generated by the Type-I seesaw and not by some  extended seesaw models like the inverse or linear ones, where the relevant couplings could be orders of magnitude larger.

Clearly, the requirement introduced by the CM energy of the ILC being fixed at discrete values all below the TeV scale, as opposed to the LHC case where, at the partonic level, $\sqrt{\hat{s}}$ can be well beyond it, limits the region of NMSSMr parameter space that can be accessed. However, within this part of parameter space, we can define representative BPs which show that we can see a signal from Yukawa couplings smaller than $10^{-6}$. 

\vspace*{0.25cm}\noindent
{\bf Acknowledgments} SM is financed in part through the NExT Institute and the STFC consolidated Grant No.
ST/L000296/1. HW acknowledges financial support from the Finnish Academy of Sciences and
Letters and STFC Rutherford International Fellowship scheme (funded through the MSCA-
COFUND-FP Grant No. 665593). The authors acknowledge the use of the IRIDIS High Performance Computing Facility and associated support services at the University of Southampton, in the completion of this work.

\end{document}